\documentclass{aa}
\newcommand{\bit}{\begin{itemize}}
\newcommand{\eit}{\end{itemize}}
\newcommand{\ben}{\begin{enumerate}}
\newcommand{\een}{\end{enumerate}}
\newcommand{\bde}{\begin{description}}
\newcommand{\ede}{\end{description}}

\usepackage{graphicx}
\usepackage{lscape}
\usepackage{natbib}
\bibpunct{(}{)}{;}{a}{}{,}

\begin{document}
\title{Simulations of galactic winds and starbursts in galaxy clusters}
\author{W. Kapferer$^1$,
        C. Ferrari$^1$,
        W. Domainko$^1$,
        M. Mair$^1$,
        T. Kronberger$^{1,2}$,
        S. Schindler$^1$,
        S. Kimeswenger$^1$,
        E. van Kampen$^1$,
        D. Breitschwerdt$^3$ \&
        M. Ruffert$^4$}

\institute{ $^1$Institut f\"ur Astrophysik,
            Leopold-Franzens-Universit\"at Innsbruck,
            Technikerstr. 25,
            A-6020 Innsbruck, Austria\\
            $^2$Institut f\"ur Astrophysik,
            Universit\"at G\"ottingen,
            Friedrich-Hund-Platz 1,
            D-37077 G\"ottingen, Germany\\
            $^3$Institut f\"ur Astronomie,
            T\"urkenschanzstrasse 17,
            A-1180 Wien, Austria\\
            $^4$School of Mathematics,
            University of Edinburgh,
            Edinburgh EH9 3JZ, Scotland, UK}

\offprints{W. Kapferer, \email{wolfgang.e.kapferer@uibk.ac.at}}

\date{-/-}

\abstract{ We present an investigation of the metal enrichment of
the intra-cluster medium (ICM) by galactic winds and merger-driven
starbursts. We use combined N-body/hydrodynamic simulations with a
semi-numerical galaxy formation model. The mass loss by galactic
winds is obtained by calculating transonic solutions of steady
state outflows, driven by thermal, cosmic ray and MHD wave
pressure. The inhomogeneities in the metal distribution caused by
these processes are an ideal tool to reveal the dynamical state of
a galaxy cluster. We present surface brightness, X-ray emission
weighted temperature and metal maps of our model clusters as they
would be observed by X-ray telescopes like XMM-Newton. We show
that X-ray weighted metal maps distinguish between pre- or
post-merger galaxy clusters by comparing the metallicity
distribution with the galaxy-density distribution: pre-mergers
have a metallicity gap between the subclusters, post-mergers a
high metallicity between subclusters. We apply our approach to two
observed galaxy clusters, Abell 3528 and Abell 3921, to show
whether they are pre- or post-merging systems. The survival time
of the inhomogeneities in the metallicity distribution found in
our simulations is up to several Gyr. We show that galactic winds
and merger-driven starbursts enrich the ICM very efficiently after
z=1 in the central ($\sim$ 3 Mpc radius) region of a galaxy
cluster.

 }
\authorrunning {W. Kapferer et al.}
\titlerunning {Galactic winds and starbursts in simulated galaxy clusters}
\maketitle

%

\section{Introduction}

Modern X-ray observations of galaxy clusters show clearly a
non-uniform, non-spherical distribution of metals in the ICM (e.g.
Schmidt et al. 2002; Furusho et al. 2003; Sanders et al. 2004;
Fukazawa et al. 2004; Hayakawa et al. 2004). As heavy elements are
only produced in stars the processed material must have been
ejected into the intra-cluster medium (ICM) by cluster galaxies.
The first suggested transfer-processes were galactic winds (De
Young 1978) and ram-pressure stripping (Gunn \& Gott 1972). Other
processes like kinetic mass redistribution due to galaxy-galaxy
interactions (Kapferer et al. 2005; Gnedin 1998), intra-cluster
supernovae (Domainko et al. 2004) or jets of AGNs are the latest
suggestions for enriching the ICM. In order to distinguish between
the efficiency of the enrichment processes several approaches for
simulations were carried out. De Lucia et al. (2004) used combined
N-body and semi-analytical techniques to model the intergalactic
and intracluster chemical enrichment due to galactic winds.
Another approach are Tree+SPH simulations of galaxy clusters
(Tornatore et al. 2004) including galactic winds. Gnedin (1998)
did combined softened lagrangian hydrodynamic (SLH)
particle-particle/particle-mesh (P$^3$M) simulations to
investigate the contribution of galaxy-interactions on the
enrichment of the ICM. A comparison of the efficiency between
ram-pressure stripping and quiet
galactic winds was recently done by Schindler et al. (2005).\\
Multiwavelength and spectroscopic observations of galaxy clusters
help us to reveal the dynamical and evolutionary state of the
system (Ferrari et al. 2005, Belsole et al. 2005). If there is no
spectroscopic information for cluster galaxies available,
questions on substructures in galaxy clusters and their dynamical
state are almost impossible to address. In this paper we show the
possibilities of state of the art and planned future X-ray
observations to obtain information about the dynamical state of a
galaxy cluster. We model galactic winds and starbursts due to
galaxy mergers as enrichment processes of the ICM. The resulting
X-ray surface brightness, temperature and metal maps provide the
key to understand the dynamics of merging and relaxed galaxy
clusters.

\section{Simulations}

\subsection{Numerical methods}

\begin{figure}
\begin{center}
{\includegraphics[width=8.8cm]{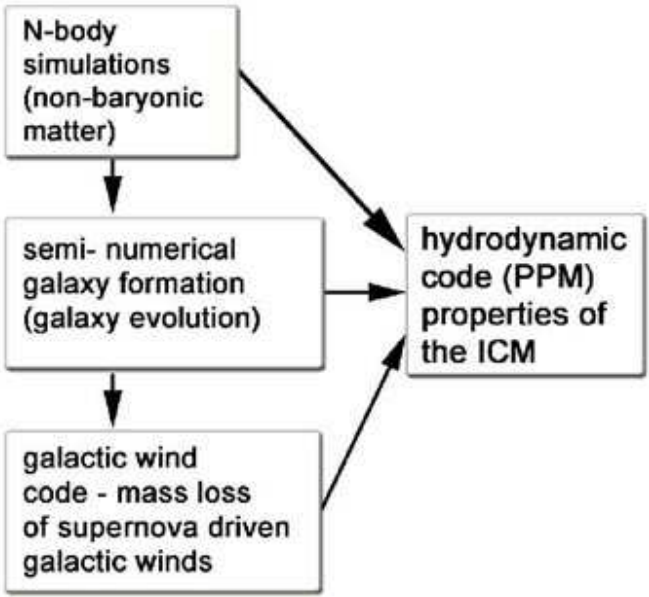}} \caption{Sketch of
our computational scheme.} \label{sketch-scheme}
\end{center}
\end{figure}

We use different code modules to calculate each of the cluster
components appropriately. Figure \ref{sketch-scheme} gives a short
overview of our involved techniques to simulate the several
components of a galaxy cluster.

\begin{figure}
\begin{center}
{\includegraphics[width=8.8cm]{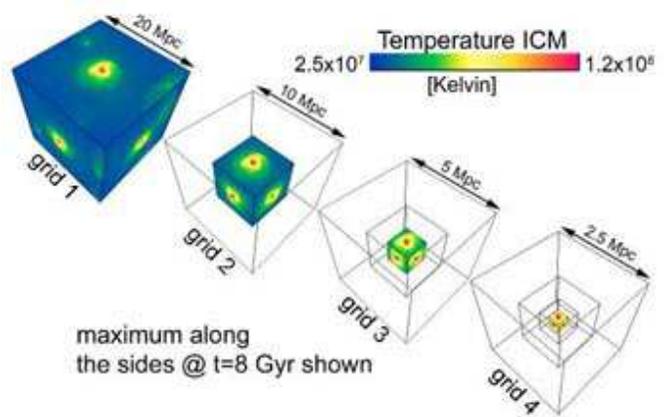}} \caption{The four
nested grids in our mesh refinement hierarchy. The largest grid
spans a volume of (20 Mpc)$^3$ the finer grids (10 Mpc)$^3$, (5
Mpc)$^3$ and (2.5 Mpc)$^3$. Each grid has a resolution of 128$^3$
grid cells. This leads to a resolution of ($\sim$19.5 kpc)$^3$ in
the innermost grid.} \label{nested_grids}
\end{center}
\end{figure}

\noindent The non-baryonic component is calculated using an N-body
tree code (Barnes \& Hutt 1986) with constrained random fields as
initial conditions (Hoffman \& Ribak 1991), implemented by van de
Weygaert \& Bertschinger (1996). The N-body tree code provides the
underlying evolution of the dark matter (DM) potential for the
hydrodynamic code and the orbits of the model cluster galaxies.
The properties of the galaxies are calculated by an improved
version of the galaxy formation code of van Kampen et al. (1999).
The adopted $\Lambda$CDM cosmology is characterized by
$\Omega_\Lambda$=0.7, $\Omega_m$=0.3, $\sigma_8$=0.93 and $h$=0.7.
The mass resolution of our N-body simulation is in the order of
$1.5 \times 10^{10}$ M$_{\odot}$ for all three model clusters with
a gravitational softening of 14 $h^{-1}$ kpc. The improvements to
the galaxy formation model of van Kampen et al. (1999) mainly
concern the star formation modes, which are a mixture of bursting
and quiescent star formation. Most of the recent star formation is
occurring in disks, following the Schmidt law with a threshold
according to the Kennicutt criterion, and most of the
high-redshift star formation is resulting from merger-driven
starbursts. The chemical evolution of stars and both the hot and
cold components are traced, where metals are assumed to be
exchanged between these components due to cooling, star formation,
and feedback processes. The galaxy formation model does not eject
metals from the galaxies. This is treated by the hydrodynamical
code. The dark matter haloes are populated with galaxies according
to the halo and galaxy merger prescriptions of the galaxy
formation model, which traces their merger histories 'ab initio'.
For each galaxy its position and velocity are known (and stored)
at all times: their orbits, luminosities, colours, etc. are a
product of the N-body simulation in combination with the
assumptions of the galaxy formation model. Dark matter haloes are
identified using local density percolation, as described fully in
van Kampen (1997). The galaxy formation code takes into account
heating of cold gas in the inter-stellar medium (ISM) due to
supernovae and enriching the hot gas in the galaxies' halos. The
cooling of the supernova (SN) heated gas is modelled by radiative
cooling with line emission. Besides the process of radiative
cooling, the hot halo gas can be reheated up to the virial
temperature by DM halo merging. This leads to a two-phase gas
model for each galaxy. A hot gas phase populating the halos and a
cold gas phase populating the disks of the galaxies. The whole gas
is gravitationally bound to the potential field of the galaxies.
Due to stellar feedback the hot gas is chemically enriched as the
systems evolve, see section 2.3 and van Kampen et al. (1999) for
details. The fraction of mass residing in the hot and cold phases
of gas depends on the SN rates and the energetics, i.e. masses and
velocities, of the DM halo mergers.\\
For the treatment of the ICM we use a hydrodynamic code with shock
capturing scheme (PPM, Collela \& Woodward 1984), with a fixed
mesh refinement scheme (Ruffert 1992) on four levels and radiative
cooling. In Fig. \ref{nested_grids} the scheme of the fixed mesh
refinement is shown.

\begin{figure}
\begin{center}
{\includegraphics[width=8.8cm]{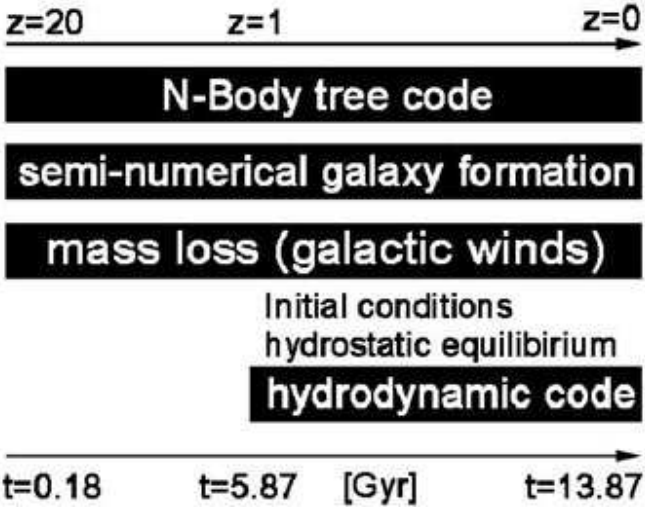}} \caption{The
chronology of our simulation set-up. While the N-body code, the
galaxy formation code and the galactic winds code trace the whole
evolution from the beginning the hydrodynamic covers $\sim$ 57\%
of the simulation time.} \label{time_scheme}
\end{center}
\end{figure}

\noindent The grids are fixed at the centre of the simulated
galaxy cluster and the largest grid covers a volume of (20
Mpc)$^3$. Each finer grid covers $\frac{1}{8}$ of the next larger
grid. With a resolution of 128$^3$ grid cells in each grid we
obtain a resolution of ($\sim$19.5 kpc)$^3$ for each cell on the
finest grid. The interaction of the N-body tree code with the
hydrodynamic code is done as illustrated in Fig.
\ref{time_scheme}.

\noindent As the N-body tree code, the semi-numerical galaxy
formation and evolution and the mass-loss rates due to galactic
winds are calculated from the beginning, the hydrodynamics covers
the redshift interval from z=1 to z=0, i.e. $\sim$ 57\% of the
whole simulation time. This setup allows us to address questions
like:

\begin{itemize}
\item{what is (are) the origin(s) of the inhomogeneities found in
the X-ray metallicity maps?} \item{how long can these
inhomogeneities survive in the hot ambient ICM?} \item{what can we
learn about the dynamical state of a galaxy cluster from X-ray
observations?}
\end{itemize}

\subsection{Calculation of the mass loss due to galactic winds}

\begin{figure}
\begin{center}
{\includegraphics[width=8.8cm]{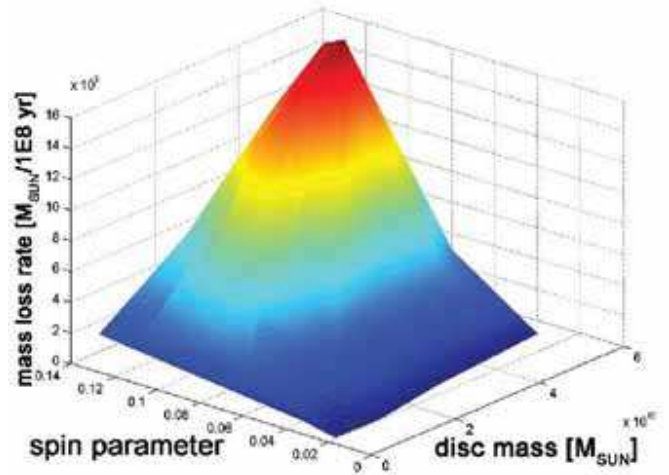}} \caption{The mass
loss rate [M$_{\sun }$/ 1$\times10^{8}$ yr] for simulated galaxies
as a function of disk mass and spin parameter ($\lambda$). The
halo mass for all galaxies is fixed to 3$\times 10^{11}$
M$_{\sun}$. It can clearly be seen that higher disk masses and
spin parameter result in a higher mass loss rate. This is due to
the fact that higher spin parameters cause a larger disk scale
length and therefore a larger disk.} \label{masslosssurface}
\end{center}
\end{figure}

The mass loss rate due to thermally, cosmic ray and MHD wave
pressure driven galactic winds is calculated for each model galaxy
with a code developed by Breitschwerdt et al. (1991). For a given
model galaxy the algorithm calculates the mass-loss rate and wind
properties as well as the velocity of the ejected matter as a
function of distance to the galaxy or the pressure flow. As input
for the wind code, galaxy parameters like halo mass, disk mass,
spin parameter, scale length of the components, temperature and
density distribution of the ISM, as well as the stellar density
distribution are required. In order to save computing time we
performed parameter studies. We did more than 10.000 calculations
varying the halo mass, disk mass and spin parameter to simulate a
wide spectrum of different spiral galaxies. The winds were then
calculated as suggested in Breitschwerdt et al. (1991). The
results are summarised in a look up table. In Fig.
\ref{masslosssurface} an example for a given set of galaxies is
shown. All galaxies have the same halo mass of 3$\times10^{11}$
M$_{\sun}$. The x- and y-axis indicate different disk masses and
spin parameters $\lambda$. The combination of disk mass and spin
parameter results then in a certain geometry (disk scale length)
and potential (Mo et al. 1998), which then is taken as input for
the galactic wind code of Breitschwerdt et al. (1991). If for a
given halo mass the fraction of disk mass and the spin parameter
increases, the size of the disk gets larger and therefore
supernovae can explode at larger radii. The gas disk as well as
the star distribution extends into the outer parts of a galaxy,
where the potential is more shallow than in the inner parts.
Supernovae exploding in that region can more easily accelerate the
surrounding gas to the escape velocity of the system at a given
galactic radius.
\noindent It is important to point out, that the mass, calculated
with the above mentioned approach, is no longer gravitationally
bound to the system. The mass is accelerated up to the escape
speed and will leave the hot halo of the underlying galaxy. Apart
from the mass loss due to galactic winds another solution is in
principle possible, the so called "breeze solution". In this
picture gas will be accelerated to leave the galactic disk, and
become then part of the hot halo gas. In this case the gas can
cool radiatively and sink then towards the disk. In our
hydrodynamic simulation only the ICM is modelled ($\rm{several}\;
10^7 < \rm{T_{ICM}} < \rm{several}\;10^8$ [K]), therefore hot gas
in the halos of galaxies is taken into account, as it is not part
of the ICM. This means that the mass expelled by supersonic
galactic winds does not belong to the hot halo component of the
galaxies anymore and is therefore part of the ICM, which we treat
in our hydrodynamic simulation.

\subsection{Calculation of the mass loss due to starbursts}

The mass loss rates due to galaxy merger induced starbursts are
estimated following Heckmann et al. (2000). They investigated 32
far-IR-bright starburst galaxies and conclude that for galactic
superwinds (mass loss rate 10-100 M$_{\sun}$ per year) the mass
outflows are comparable to the star formation rates (SFRs) of the
underlying system. The SFRs in merging galaxies in our
semi-numerical galaxy model are enhanced up to several 100
M$_{\sun}$/yr, depending on the relative velocities, masses and
the gas content of the interacting galaxies during several 10
million years. In addition, Heckmann (2003) finds that in none of
the cases radiative cooling of the coronal gas is sufficient to
quench the outflow. This heuristic approach gives us the mass loss
for a starbursting galaxy, which we add to the hydrodynamic
simulation at the place of the starburst. The material is then
spread over volumes corresponding to a typical travel time of
$10^7$ years (typical time step). The metals are mixed with the
ICM already present. The metallicity of the ejecta is taken from
the semi-numerical galaxy modelling, which calculates the metals
in a hot and cold gas reservoir through

\begin{eqnarray}
\dot{m_c} &=& Z_h \dot{M}_{cool}-Z_c(1+ \beta -R) \Psi_{SF}+y
\Psi_{SF}\\
\dot{m_h} &=& Z_h \dot{M}_{cool}+Z_c \beta \Psi_{SF}+Z_{Prim}
\dot{M}_{new},
\end{eqnarray}
\noindent where $Z_c=m_c/M_c$, $Z_h=m_c/M_h$, $\dot{M}_{new}$ is
the accretion of new baryons by the halo, $\Psi_{SF}$ is the star
formation rate, $Z_{prim}$ the primordial metallicity and $\beta$
the fraction of supernovae of a stellar population. The starbursts
are modelled such, that they eject matter with the metallicity
present in the cold gas reservoir. This results in a lower limit
in metallicity of the starburst ejecta.

\section{The properties of our model galaxy clusters}

Three different model galaxy clusters have been simulated,
hereafter cluster A, B and C. We simulate a non-merging galaxy
cluster, a cluster with a single subcluster merger and a less
massive galaxy cluster undergoing several mergers of subclusters.
In Table \ref{galaxy cluster properties} the properties of the
model clusters A, B and C are listed.

\begin{table}
\begin{center}
\caption[]{Properties of our model galaxy clusters}
\begin{tabular}{c c c}
\hline \hline & total mass inside & \# of galaxies  \cr model &  a
radius of 3 Mpc & in the inner (5 Mpc)$^3$\cr cluster & @ z=0
[10$^{14}$M$_{\sun}$] & volume (grid 3) \cr\hline A & 13.0 & 830
\cr B & 6.8 & 489 \cr C & 7.8
 & 585 \cr\hline
\end{tabular}
\label{galaxy cluster properties}
\end{center}
\end{table}

\begin{figure}
\begin{center}
{\includegraphics[width=8.8cm]{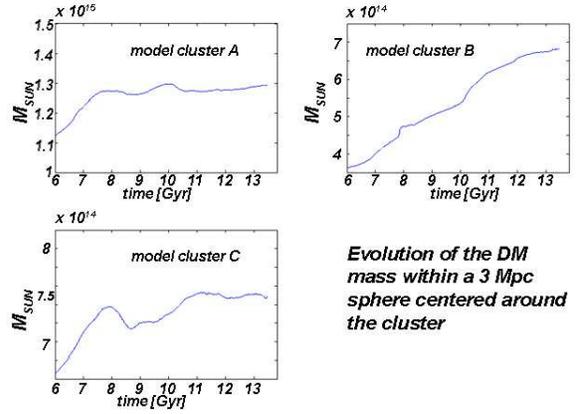}} \caption{From the
mass accretion rates it is visible that model cluster B is a
strong merging system, model cluster C has less merging events and
model cluster A is a relaxed, massive system. The Evolution is
shown for a 3 Mpc physical units sphere centered around the
cluster, since z=1.} \label{massaccretion}
\end{center}
\end{figure}

\begin{figure}
\begin{center}
{\includegraphics[width=8.8cm]{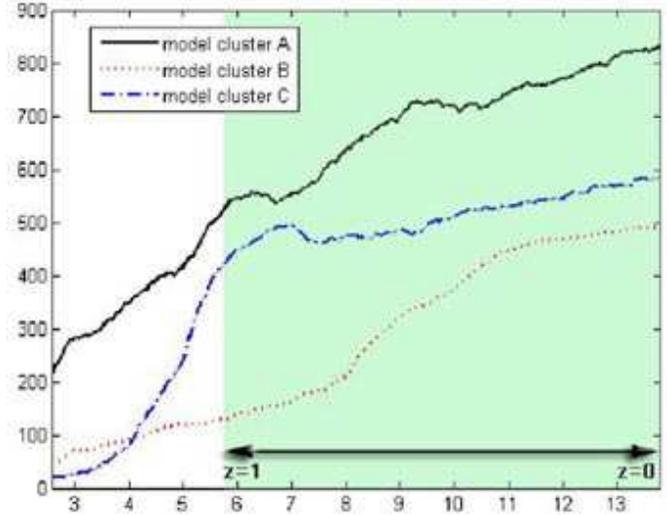}} \caption{Time
evolution of the number of galaxies in the simulation volume grid
3 (5 Mpc on a side) since z=4. The blue range highlights the
redshift interval z=1 to z=0.} \label{evolution_all galaxies}
\end{center}
\end{figure}

\noindent In Fig. \ref{massaccretion} the evolution of the DM mass
within a 3 Mpc sphere centered around the cluster is shown. The
number of galaxies in our model clusters as a function of time in
the simulation volume of grid 3 (5 Mpc on the side) is shown in
Fig. \ref{evolution_all galaxies}.

\begin{figure}
\begin{center}
{\includegraphics[width=8.8cm]{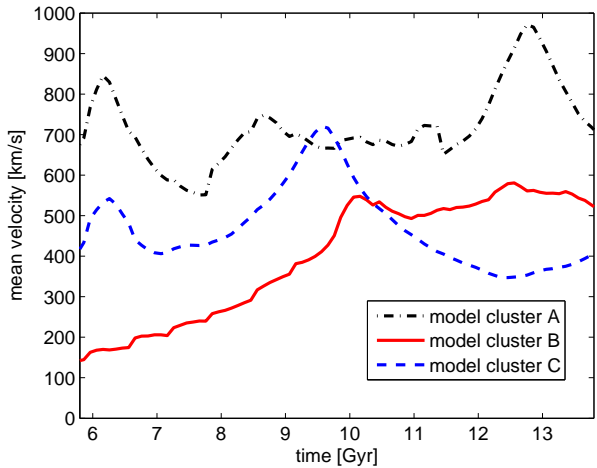}} \caption{Time
evolution of the mean velocity of all galaxies in a central cube
with 5 Mpc on the side for model cluster A, B and C. The
velocities are given starting from z=1 (blue range in Fig.
\ref{evolution_all galaxies}.)} \label{galaxies_velo}
\end{center}
\end{figure}

\noindent As model cluster B is the cluster with several major
mergers, the number of galaxies in grid 3 (5 Mpc per side) has a
stronger increase (increase = new galaxies are formed + galaxies
move into the volume) with time than in the cases of the
non-merging model cluster A, and less merging system C, which have
an almost constant number of galaxies in the same volume. The
average mass ratio of the subclusters and mainclusters in our
simulations is around 1:2 in model cluster B, and 1:4 in model
cluster C. In model cluster A, a very low mass subcluster falls
towards the cluster centre almost at the end of the simulation.
Note that the pre-merger clusters have a lot of substructure. The
mean velocities of the galaxies as a function of time are shown in
Fig. \ref{galaxies_velo}. Model cluster A has a approximately
constant mean velocity over a long period and a little merger
event during 1.5 Gyr at the end of the simulation. Model cluster C
has two major merger events, one at the beginning and one in the
middle of the hydrodynamic simulation; see the two humps in the
corresponding line in Fig. \ref{galaxies_velo}. The less massive
model cluster B is forming during the first half of the
hydrodynamic simulation by mergers of several small substructures
and then in the second half there are two major mergers, at 10 and
12.5 Gyr.
\noindent In our model clusters the mass loss rate due to both
quiet galactic winds and starbursts are calculated. The ratio of
matter ejected by all starburst events in comparison to the
galactic winds in the time interval z=1 to z=0 is 1.3\% for model
cluster A,  20\% for model cluster B and 7.3\% for model cluster C
in a volume of (5 Mpc)$^3$ centred on the cluster. As model
cluster B shows the highest number of merging events, it has the
highest fraction of starburst events in the (5 Mpc)$^3$ simulation
volume. This is caused by many galaxy groups moving towards the
cluster centre as the cluster evolves.

\section{Results}
\subsection{X-ray maps}

\begin{figure*}
\begin{center}
{\includegraphics[width=\textwidth]{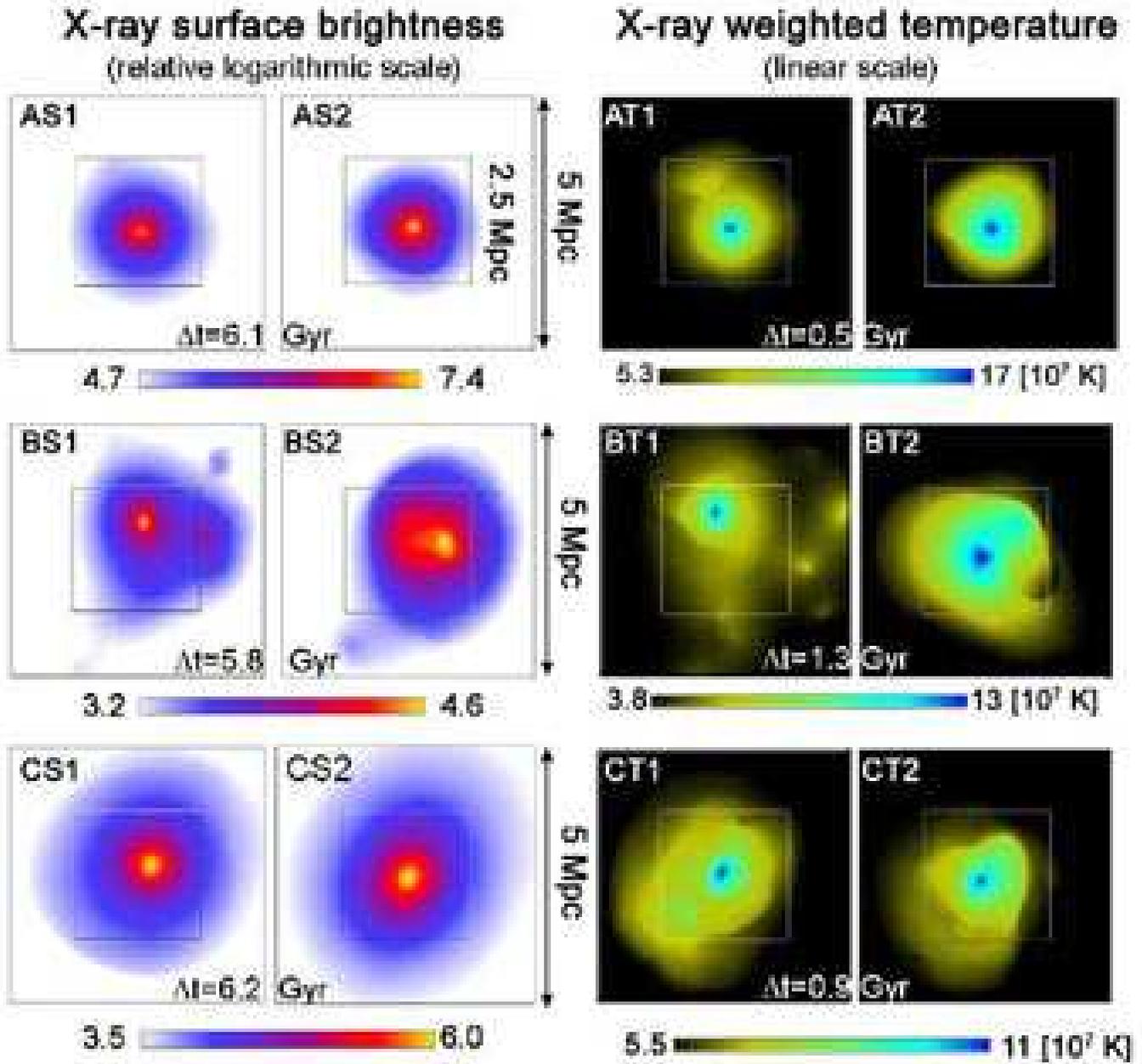}} \caption{X-ray
surface brightness maps and X-ray weighted temperature maps for
all model clusters. In the upper panel model cluster A, in the
middle panel model cluster B and in the lower panel model cluster
C is shown. The name for each map indicates the model cluster (A,B
or C), surface brightness or temperature map (S, T) and two
particular timesteps 1 or 2. The timesteps are chosen to show the
clusters before and after a merger event. The X-ray surface
brightness and temperature maps correspond not to the same
timesteps, the main criterion for choosing them is the visibility
of substructures. The range of the surface brightness maps is
about 2.5 orders of magnitudes. See the text for more details.}
\label{surface-temperature}
\end{center}
\end{figure*}

We extract X-ray surface brightness, X-ray emission weighted
temperature and metal maps from our simulations for comparison
with observations. The images in Fig. \ref{surface-temperature}
and \ref{metal-maps} have a resolution of 256 pixels on each side
(one pixel represents a region of (19.5 kpc)$^2$ and represent the
projection of a volume of (5 Mpc)$^3$ on the sky).

\subsubsection{X-ray surface brightness maps}

Fig. \ref{surface-temperature} shows all model clusters A, B and C
at different evolutionary states. Each panel on the left hand side
shows two X-ray surface brightness maps (named S1 and S2),
reproducing the X-ray emission at two different particular states
in their evolution. The first letter of the name indicates the
model cluster A, B or C. The maps AS1 and AS2 (upper panel) in
this series show model cluster A, the most massive one with an
infall of a low mass structure happening almost at the end of the
simulation. In AS1 there is a small area with enhanced emission in
the innermost part (2.5 Mpc)$^2$ faintly visible at the top. This
feature belongs to a small subcluster. At the end of the
simulation the whole cluster has become a relaxed system.\\
The maps BS1 and BS2 demonstrate the different dynamical states of
model cluster B. At z=1 (map BS1) there are two subclusters
visible, besides the main cluster. In addition there is enhanced
X-ray emission due to matter ejected by a major starburst in the
upper right corner. The overdensities in the surroundings of the
starbursting galaxy cause the conspicuous behaviour in the X-ray
surface brightness map. Map BS2 shows the state when the two
substructures have already merged with the main cluster. The
ejected matter due to the starburst, caused by the galaxy merger,
has moved to the vicinity of the centre and is visible as a second
small local maximum of X-ray emission. The elongated shape in the
X-ray surface brightness map due to the merger is clearly visible.
The cluster is in a post merger state.\\
The X-ray surface brightness map for model cluster C is shown in
maps CS1 and CS2. The first map CS1 shows the cluster at the
beginning of the hydrodynamic simulation. As the cluster undergoes
one major merger event at z=0.5, the system does not evolve to a
relaxed galaxy cluster within simulation time. In map CS2 the
X-ray surface brightness map at z=0 is shown. The elongated
structure, caused by the merger, can clearly be seen.\\
X-ray surface brightness maps are able to show major merger events
in galaxy clusters. X-ray observations are often limited by the
number of photons arriving from the observed object and of course
the intrinsic noise. Therefore it is important to mention that the
maps that are shown here represent an ideal case of infinitely
long observations.

\subsubsection{X-ray temperature maps}

In Fig. \ref{surface-temperature} the right images (named T1 or
T2) show the X-ray weighted temperature maps of all model clusters
at the same time as the same time as the left panels. Again a
volume of (5 Mpc)$^{3}$ is represented by the maps. The two maps
AT1 and AT2 show model cluster A before a small merging event and
during the phase of merging. Map AT1 shows the cluster at the same
time as AS1. In the X-ray temperature map the subcluster is more
pronounced and the map shows in addition more structures in the
centre of the main cluster, caused by several subgroups of
galaxies falling in. Map AT2 shows the cluster during the merging
phase 500 Myr after map AT1. The subcluster has fallen into the
main cluster and caused perturbations in the ICM, which can be
seen in the aspherical shape of the cluster's temperature
distribution.\\
Model cluster B undergoes several merging events of subclusters
which can be seen in the map BT1. Besides the subclusters, map BT1
shows small, colder regions (black knots) in the ICM, which were
caused by cold matter ejected by starbursts. They can be seen in
the outskirts of the main cluster. In map BT2 the cluster is shown
after 1.3 Gyr of evolution. The subclusters have merged with the
main cluster and strong shocks in the ICM emerge, which are
visible near the upper border of the indicated (2.5 Mpc)$^2$
region. In the right upper corner of the (2.5 Mpc)$^2$ cube a cold
feature is visible. This feature belongs to ejected matter by a
merger-driven starburst. The ejected matter has several 10$^6$ K
and is therefore visible as cold inclusions in the ICM. The
temperature of the ICM in the region of the starburst is
calculated as a mixture of hot ICM with a given density and the
ejecta of the starburst, which are assumed to have temperatures in
the order of several 10$^6$ K. As starbursts can transport a major
amount of matter into the ICM, the density increases and therefore
the regions are visible in X-ray maps, see Fig.
\ref{surface-temperature} - BS1.\\
The last model cluster C has a major merger event, as is clearly
visible in the X-ray weighted temperature map. Map CT1 shows the
subcluster and in addition cold material on the opposite side of
the main cluster, which belongs to ejected matter from a
starburst. In the centre of the main cluster there are several
outward moving shocks showing up as hotter shell-like regions,
which were caused by several very small subcluster mergers. In map
CT2 the subcluster has merged with the main cluster (900 Myr
later).\\
X-ray weighted temperature maps are better suited to investigate
the dynamical state of a galaxy cluster than the X-ray surface
brightness maps. If enough photons are available in the X-ray
observations, combination of both gives the best tool for
investigating galaxy clusters in X-rays. The observed X-ray
temperature maps of galaxy cluster have a limited spatial
resolution, because the numbers of photons in real observations is
limited.

\subsubsection{X-ray metal maps}

X-ray weighted metal maps show the strength of enrichment
processes and the spatial spreading of metals. As we focus in this
paper on the enrichment of the ICM by galactic winds and
starbursts, the metal maps give a direct measure of their
efficiency.
\begin{figure}
\begin{center}
{\includegraphics[width=\columnwidth]{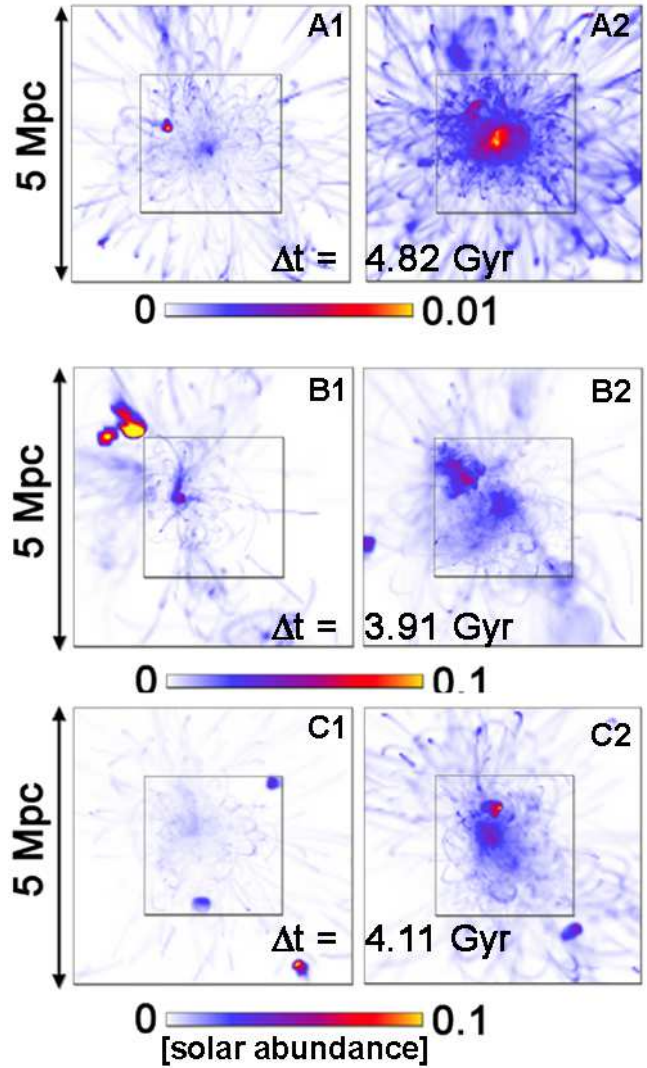}}
\caption{X-ray metallicity maps for all model clusters. The names
of the images reflect the model cluster (A, B or C) shown at 2
different timesteps. See the text for more details. Between A1 and
A2 there are 4.82 Gyr, between B1 and B2 are 3.91 Gyr and between
C1 and C2 4.11 Gyr of evolution.} \label{metal-maps}
\end{center}
\end{figure}
\noindent In Fig. \ref{metal-maps}, upper panels, two metal maps
for model cluster A (A1, A2) are shown. The main striking
features, which are visible, are stripes of enriched material
pointing radially toward the cluster centre. These stripes belong
to enriched matter ejected continuously by quiet galactic winds as
the galaxies move through the ICM. As we calculate the amount of
ejected matter by SN-driven galactic winds, the enriched gas is
treated in the same hydrodynamic way as the ICM. In the outskirts
of the galaxy cluster the stripes can survive several Gyr because
the density and strength of perturbations of the ICM in enriched
regions is very low. In section 4.5 a more detailed analysis of
the survival timescales will be given. A1 and A2 are separated by
4.82 Gyr of evolution. In A1 there is a small highly enriched
region visible, which belongs to matter ejected by a merger-driven
starburst. As model cluster A is a very massive galaxy cluster
(see Table \ref{galaxy cluster properties}) the pressure of the
ICM on the galaxies is quite high in the central region of the
cluster. Suppression of winds occurs, because enhanced $pdV$-work
against the increasing ambient pressure can stall the expansion of
the outflow. In our model this translates into a combined pressure
of gas, cosmic rays and MHD waves, which is lower at the sonic
point than the external ICM pressure. In addition to the quiescent
winds, calculated in Schindler et al. (2005), we take into account
starbursts caused by galaxy-galaxy interactions in this
simulation. These interactions enhance the SFRs of the systems
very strongly over short times (several 10 to 100 Myr, Kapferer et
al. 2005) and therefore lead to short, strong outflows of matter.
In A2 the highly enriched central region can be seen. The overall
metallicity is very low in model cluster A, because of the
suppression of quiet galactic winds by ICM pressure onto the
galaxies in the central part ($\sim$ (1 Mpc)$^3$). Another
interesting feature is the second peak in the metal map near the
cluster centre (near upper-left inner corner) in the map A2 in
Fig. \ref{metal-maps}, caused by the strong starburst from above.
The material has not been mixed with the metals at the centre of
the main cluster. It has been mixed with gas in a volume of
several tens of kpc. As shown in the previous maps, model cluster
A has a merger event and the induced perturbations in the ICM
cause the spatial shift of this enriched region to the
upper left corner of the innermost grid.\\
In Fig. \ref{metal-maps} B1 and B2, the X-ray metallicity maps of
model cluster B are shown. The left map shows the metallicity
distribution after 4 Gyr of evolution. Again two highly enriched
regions in the outskirts of the cluster are visible. They belong
to merger-driven starburst ejecta. Note that the relative
velocities of galaxies, which fall in groups toward the galaxy
cluster centre are significantly lower (several hundred km/s) than
the relative velocities of the galaxies interacting at the centre
of the cluster potential (thousands of km/s). As a consequence,
the interaction times of galaxies in subgroups falling into the
cluster potential are longer and merger events do have larger
timescales to introduce perturbations into the inter-stellar
medium, which cause the higher SFRs. Therefore starbursts due to
galaxy-galaxy interactions are more effective in subgroups,
falling into the cluster potential, than in the central parts of
the galaxy cluster. In the metal B2 map the ICM, enriched by the
starburst, has efficiently mixed with the surrounding ICM and
moved towards the centre of mass, i.e. into denser regions. As a
consequence the average metallicities are lower than in the left
map (B1), where the ICM was not so dense in the region, where the
starburst happened. Model cluster B is the least massive model
cluster in our sample and does not suppress galactic winds in the
central region. Although the cluster has fewer galaxies, galactic
winds and starbursts can enrich the ICM an order of magnitude more
efficiently than in model cluster A, the most massive cluster in
our sample. This can be explained in terms of the ICM mass, which
has to be enriched. A high gas mass exerts a higher pressure of
the ICM onto the galaxies, especially in the central (1 Mpc)$^3$
region. Although more massive galaxy clusters have more galaxies,
the higher pressure of the ICM suppresses the galactic winds of
many systems in the central region. In addition, more ICM has to
be enriched than in the case of a less massive cluster, which
leads to lower metal abundances in our model
cluster A.\\
In Fig. \ref{metal-maps} C1 and C2, the metal maps for the
intermediate mass model cluster C are shown. In principle they
follow the same trend as in the case of model cluster B. Again the
left map shows highly enriched regions due to galaxy-merger driven
starburst ejecta. As in all the other metal maps, the enriched
trajectories of the galaxies can be seen in the outskirts of the
galaxy cluster. These stripes can survive several Gyr, depending
on the properties and the motion of the surrounding ICM. In the
central $\sim$ (2 Mpc)$^3$ part of the model cluster, the shock
waves and perturbations in the ICM, caused by subcluster mergers,
mix the enriched regions on shorter timescales (several hundred
Myr), depending on the mass ratios and the merger scenario.

\subsubsection{Evolution of the mean metallicities}

\begin{figure}
\begin{center}
{\includegraphics[width=\columnwidth]{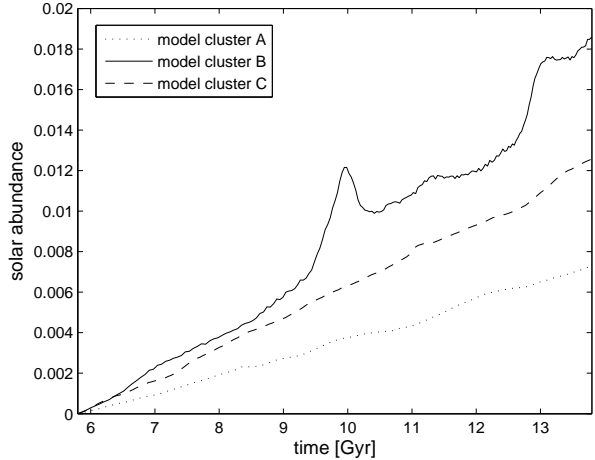}}
\caption{Evolution of the X-ray weighted metallicity for all model
clusters. The metal maps of the (2.5 Mpc)$^2$ region are weighted
by X-ray surface brightness maps in the same region.}
\label{metalevo6.26Mpc2}
\end{center}
\end{figure}

\begin{figure}
\begin{center}
{\includegraphics[width=\columnwidth]{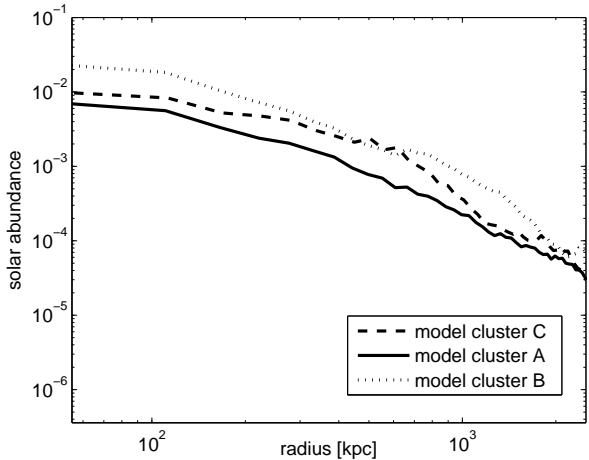}}
\caption{Metallicity profiles of the three model clusters at z=0.}
\label{profiles}
\end{center}
\end{figure}

In order to give a quantitative estimate of the enrichment, we
present the evolution of the mean metallicities of our model
clusters during the time interval z=1 to z=0 in Fig.
\ref{metalevo6.26Mpc2}. The quantities given are the mean
metallicities extracted from the X-ray weighted metal maps,
weighted with the X-ray surface brightness maps. In Fig.
\ref{metalevo6.26Mpc2} a region of (2.5 Mpc)$^2$ was taken into
account, which corresponds to a typical galaxy cluster size.
Whereas model cluster A and C show approximately a constant slope
over the simulation time, model cluster B reveals a different
behaviour. At 10 Gyr an enhancement is visible because of a merger
event, which increases the X-ray weighted metallicities due to a
temporary increase in density in the merger region. In the case of
model cluster B, the first maximum in the mean velocity (see Fig.
\ref{galaxies_velo}) corresponds to the enhancement of the
metallicity enrichment of model cluster B in Fig
\ref{metalevo6.26Mpc2}, which is related to a cluster merging
event.
\noindent In model cluster B the increase in the number of
galaxies in the simulation volume with (5 Mpc)$^3$ is stronger
than in the model clusters A and C. Therefore the increase in the
enrichment rate can be explained as increase of number of galaxies
with
galactic winds in the simulation volume.\\
In Fig. \ref{profiles} the metallicity profiles of the three model
clusters at z=0 are shown. The centre of the 2D binning coincides
in all cases with the maximum in density of the baryonic matter.

\noindent The central metallicity in model cluster B increases by
an order of magnitude within $\sim$4 Gyr, ending at z=0. Metal
profiles are easier to obtain from observations than 2D metal
maps, but they have one major disadvantage: they do not show the
inhomogeneities in the metal distribution of a galaxy cluster. The
comparison of the metal profile of model cluster B at the end of
the simulation with the corresponding artificial X-ray weighted
metal map in Fig. \ref{metalmap112_evo} shows this issue clearly.
Due to the averaging of metals in rings centred around the
emission centre of the ICM, the real maximum in the 2D metal maps
is flattened out.

\begin{figure}
\begin{center}
{\includegraphics[width=\columnwidth]{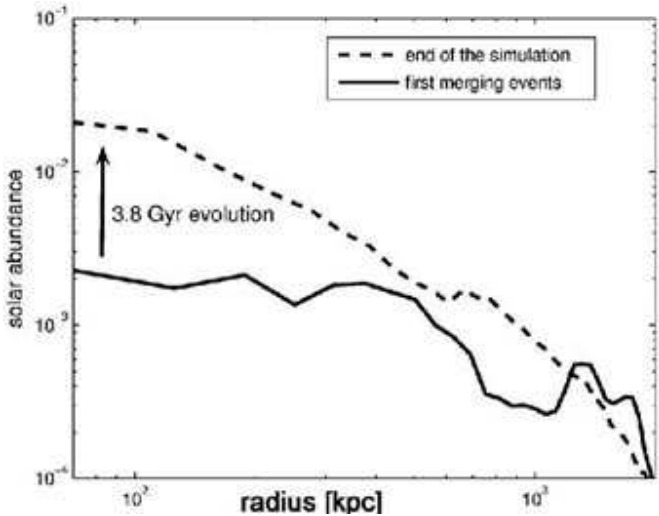}}
\caption{Metallicity profile for model cluster B at the first
merging events and at the end of the simulation.}
\label{profile112evo}
\end{center}
\end{figure}

\begin{figure}
\begin{center}
{\includegraphics[width=\columnwidth]{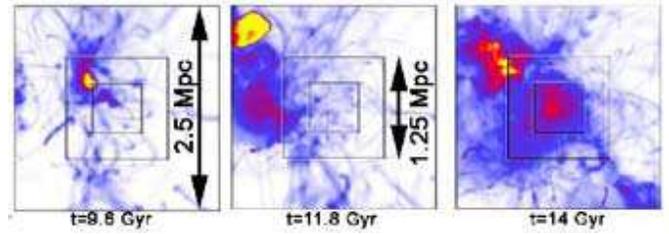}}
\caption{X-ray weighted metal maps for model cluster B at
different times. The movement of the metals with respect to the
X-ray surface brightness centre (fixed at the image centre) due to
several cluster mergers can easily be seen.}
\label{metalmap112_evo}
\end{center}
\end{figure}

\subsection{Revealing the dynamical state through X-ray maps}

\begin{figure}
\begin{center}
{\includegraphics[width=\columnwidth]{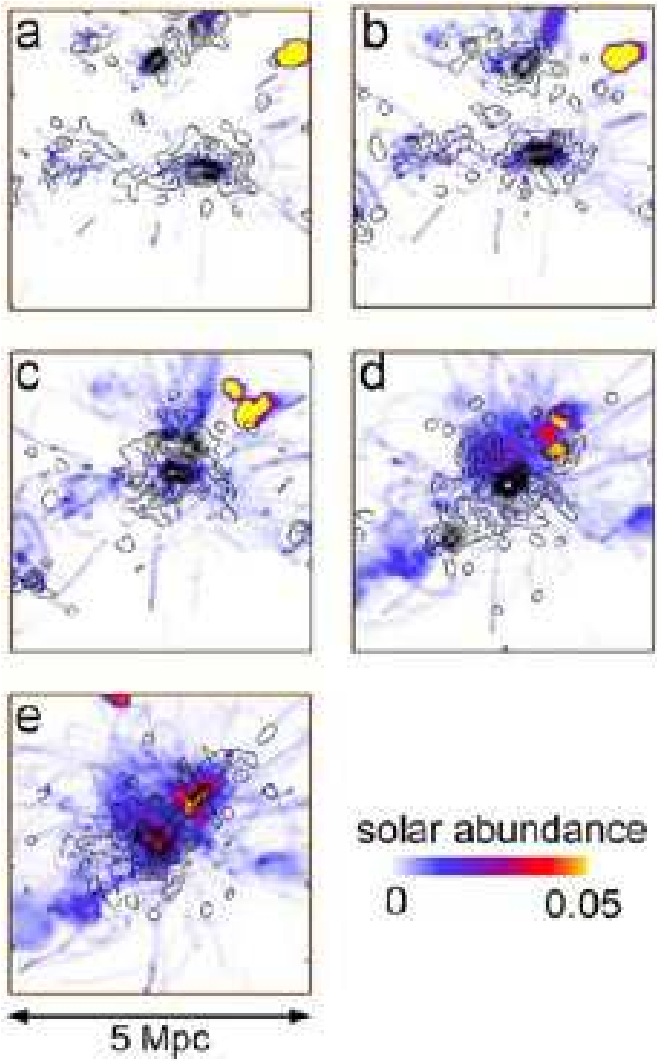}}
\caption{X-ray weighted metallicity maps of model cluster B with
galaxy-density isolines. Images (a) to (e) show the model cluster
at different times, chosen in such a way that pre- and in situ
merger evolutionary states are presented (image a to e shows about
2.5 Gyr of evolution). The metallicity-gaps in front (relative to
the direction of motion) of the substructures are clearly
visible.} \label{galaxyisolinesmetal}
\end{center}
\end{figure}

Galaxy clusters are the largest gravitationally bound systems in
the universe and, as they are formed from huge amounts of DM,
galaxies and ICM, it takes cosmological timescales to establish
dynamical equilibrium. Galaxy clusters can be observed in many
wavelengths to study the interaction of the different components.
One outstanding challenge is the comparison of high resolution
X-ray metal maps with galaxy-density maps. The combination of both
observations will lead to a deeper insight into the interaction
processes in galaxy clusters. In this section we want to present
the capabilities of these combined observations to address
questions about the dynamical state of a
galaxy cluster.\\
If metals originate from galaxies, they are expelled by processes
like ram-pressure stripping, galactic winds or direct kinetic
redistribution of material by galaxy-galaxy interaction. Therefore
they are connected to the trajectories of the galaxies. The
relative velocity of the galaxies with respect to the ICM then
leads to metal tracks, which can survive for several Gyr in the
quieter outer regions of a galaxy cluster. In Fig.
\ref{galaxyisolinesmetal} X-ray weighted metal maps for model
cluster B are shown, with isolines representing the galaxy number
density. Images (a) to (e) show different evolutionary states of
model cluster B, the model cluster with many merger events. The
images (a), (b) and (c) show the infall of a major subcluster from
the top. Image (c) shows the cluster in the pre-merger phase.
Along their way to the main cluster the galaxies lose metals and
therefore there is a higher metallicity behind the subcluster than
in front of them. In image (d) another subcluster falls into the
main cluster (from the lower left). Again there is a higher
metallicity behind the subcluster than in front of it. This
results in a gap in the metal map, between the subcluster and the
main cluster, just before the two structures merge.

\begin{figure}
\begin{center}
{\includegraphics[width=\columnwidth]{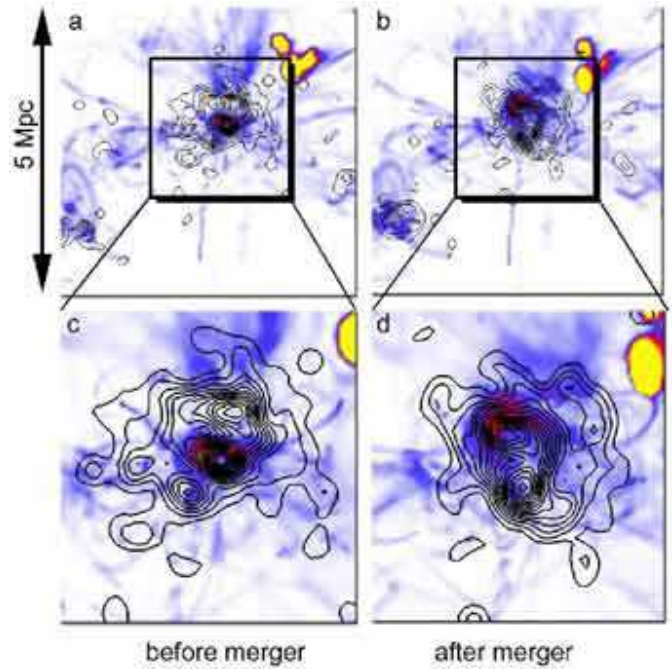}}
\caption{Artificial X-ray weighted metallicity maps of model
cluster B with galaxy-density isolines drawn in (the color bar is
the same as in Fig. \ref{galaxyisolinesmetal}) . The upper images
show again the (5 Mpc)$^2$ region, whereas the lower images are
the magnifications of the innermost grid (2.5 Mpc)$^2$. The
metallicity gap before the first encounter of the subcluster with
the main cluster is clearly visible. The main cluster is located
in the centre of the image (c) and the gap lies above.}
\label{galaxyisolinesmetalzoom}
\end{center}
\end{figure}

\noindent In Fig. \ref{galaxyisolinesmetalzoom} two timesteps of
the merging model cluster B are presented. The upper images show
the cluster shortly before ($\sim$80 Myr) and shortly after
($\sim$ 60 Myr) the first passage of the subcluster through the
main-cluster. In the lower panels (image (c) and (d))
magnifications of the central (2.5 Mpc)$^2$ cluster region are
shown. Before the first encounter (image (c)) the absence of
metals between the main- and the subcluster is clearly visible. In
image (d) the subcluster with the corresponding galaxies went
through the main cluster and therefore there is no metallicity gap
between them anymore.

\begin{figure}
\begin{center}
{\includegraphics[width=\columnwidth]{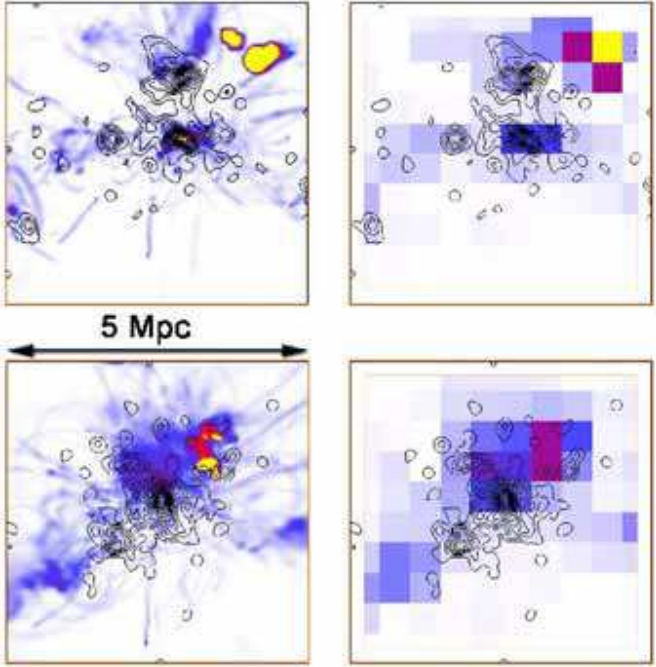}}
\caption{X-ray weighted metallicity maps of model cluster B with
galaxy-density isolines. The color bar is the same as in Fig.
\ref{galaxyisolinesmetal}. The left images show the simulation
resolution, the right images present the same metal maps with a
resolution of 9x9 pixel.} \label{galaxyisolinesmetalblocky}
\end{center}
\end{figure}

\noindent The combination of both observations (optical imaging
and X-ray) can reveal the dynamical state of a galaxy cluster very
accurately, without any time consuming optical spectroscopic
investigations. The fact that there is a strong gradient in the
metallicity distribution in front of and behind the infalling
subcluster, indicates the direction of motion of the galaxies and
the pre- or post merger state of a galaxy cluster.\\
Of course, our simulated metallicity maps have resolutions not
obtainable with present X-ray observations. Therefore Fig.
\ref{galaxyisolinesmetalblocky} shows more realistic metal maps,
including galaxy densities isolines. The right images give the
same quantities as the left ones, but with a resolution of 9x9
pixels. The metallicity gap between the main- and subcluster in
the upper images is present, in the high as well as in the low
resolution image. In the lower images another subcluster falls
into the main cluster from the lower left corner. Again the higher
metallicity behind the subcluster, which moves towards the main
cluster, is visible in both the low and high resolution metal
maps.

\subsection{Comparison with observations: the bimodal clusters
            Abell 3921 and Abell 3528}

In order to test whether the metallicity gap between pre-merging
structures in galaxy clusters are observable or not, we compared
the three model clusters with observed galaxy clusters. Two
systems in which metal gaps have been observed, and whose
dynamical state has been studied in detail through a
multi-wavelength analysis, are presented in the following: Abell
3921 and Abell 3528.

\subsubsection{Abell 3921}

\begin{figure}
\begin{center}
{\includegraphics[width=\columnwidth]{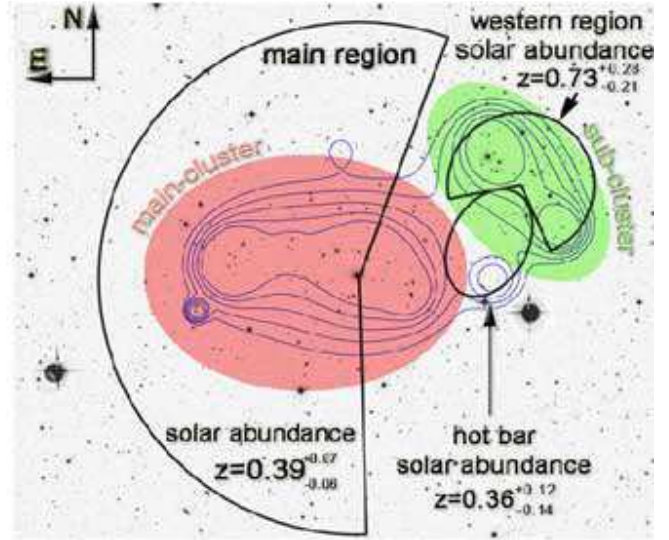}}
\caption{Iso-density contours (blue) of the projected distribution
of the red sequence galaxies (with $R_{\rm{AB}}\leq 19$)
superimposed on the central 30 $\times$ 22 arcmin$^{2}$ of the
R-band image of A3921. Metallicity of the ICM (solar abundance)
was determined from XMM data by Belsole et al. (2005) in three
regions outlined by the black lines.} \label{Abell3921}
\end{center}
\end{figure}

Abell 3921 is a R=2, BM II cluster at $z=0.094$ (Katgert et al.
1998). Recent optical and X-ray analysis (Ferrari et al. 2005,
Belsole et al. 2005) showed that its perturbed morphology, firstly
revealed by ROSAT and Ginga observations (Arnaud et al. 1996), is
due to an on-going merger of two main sub-structures: a main
cluster centred on the Brightest Cluster Galaxy (BCG) (A3921-A),
and a NW subcluster (A3921-B) hosting the second brightest cluster
member. A3921-A is $\sim$5 times more massive than A3921-B. The
comparison of the optical and X-ray properties of A3921 (i.e.
dynamical and kinematic properties of the cluster, optical and
X-ray morphology, features in the ICM density and temperature
maps) suggests that A3921-B is tangentially traversing the main
cluster in the S-SE North direction. The two sub-clusters show a
collision axis nearly perpendicular to the line of sight (Ferrari
et al. 2005, Belsole et al. 2005). \\
Belsole et al. (2005) extracted the ICM metallicity in three
different regions of A3921: the Eastern part of the main cluster
(in which the ICM is less perturbed by the merging event), the hot
bar detected on the temperature map, and the western region (see
Fig.~5 of Belsole et al. (2005) and Fig. \ref{Abell3921}). Taking
their result as best reference, the metallicity of the western
region is significantly higher ($\sim$90\% significance level)
than in the main region and in the hot bar of A3921.

\begin{figure}
\begin{center}
{\includegraphics[width=\columnwidth]{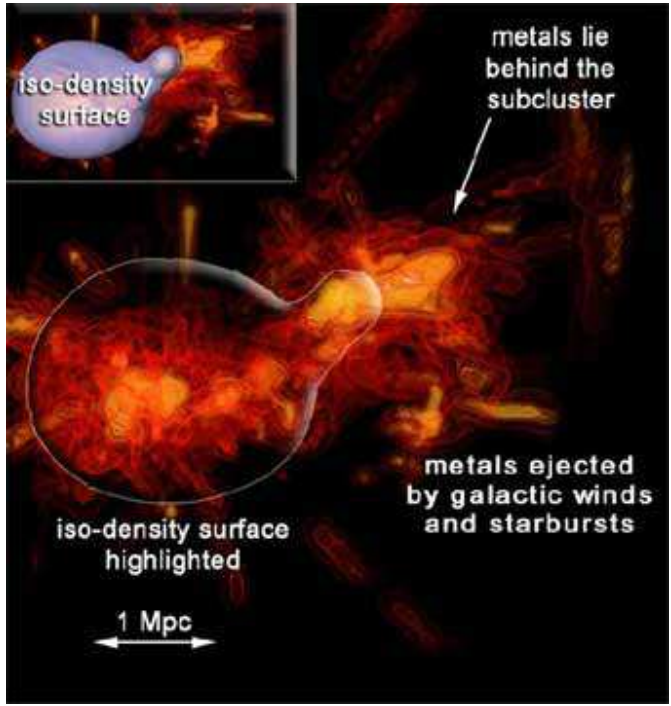}} \caption{The
3D metal distribution of model cluster B and the edge of the ICM
density isosurface (highlighted by the insert) are shown. The
subcluster has higher metallicities, due to a merger-driven
starburst, and the offset of the metals with respect to the
subcluster is visible. In addition the gap in metallicity,
indicating the pre-merger state, can be seen clearly.}
\label{metal_density_112}
\end{center}
\end{figure}

\noindent In Fig. \ref{metal_density_112} the metallicity
distribution of model cluster B is shown. The insert gives in
addition one ICM iso-density surface, which shows the positions of
the main- and the subcluster. As mentioned above, metals, ejected
by galactic winds and merger-driven starbursts, are located behind
the subcluster. As galaxies in subclusters interact with lower
relative velocities (several 100 km/s), they show stronger
enhancements in their SFRs than in the cluster central regions,
where the relative velocities are of the order of 1000 km/s. In
addition the gas content of cluster galaxies, especially in the
central regions, are lower due to ram-pressure stripping. In Abell
3921 the western region shows the highest metallicity, nearly
twice of that of the main region. If our picture holds, the
subcluster seems therefore to be in the pre-merger state.

\noindent This confirms that the previous hypothesis that A3921 is
in the phases of merging (Ferrari et al. 2005, $t_0\sim\pm$ 0.3
Gyr) is correct, but we can now even refine the merging scenario
by affirming that the closest cores encounter has not yet happened
(in partial disagreement with the conclusions of Belsole et al.
2005). If a high resolution metal map were available, we would
probably also be able to reconstruct the 3D-path of the subcluster
A3921-B.

\subsubsection{Abell 3528}

Abell 3528 is a R=1, BM II Abell cluster at $z=0.053$ located in
the core of the Shapley Concentration. X-ray observations
(Raychaudhury et al. 1991; Schindler 1996) revealed a bimodal
morphology, with two main clumps (A3528-N and A3528-S) aligned in
the North/South direction and separated by $\sim$14 arcmin (0.9
${h_{70}}^{-1}$ Mpc).  Further optical, X-ray and radio analysis
(Bardelli et al. 2001; Baldi et al. 2001; Donnelly et al. 2001;
Venturi et al. 2001) indicated that A3528 is a pre-merging system,
since: a) the galaxies in the two clumps are not far from the
virial equilibrium each, b) the merging effects on the galaxy
population are not yet evident, neither at optical nor at radio
wavelengths, c) no diffuse extended radio emission has been
detected, and d) the ICM surface brightness distribution of the
two clumps is still azimuthally symmetric, with evidences for cool
cores, and the gas located between the two clumps is only
marginally hotter ($\sim$15\%).\\
Through XMM observations of A3528, Gastaldello et al. (2003)
reached different conclusions on the dynamical state of this
cluster, assessing that in fact A3528 is an off-axis post-merging
system. They conclude that the absence of any evidence of shock
heated gas both in their temperature and in their density map is
due to a wrong initial hypothesis, i.e. the two clumps cannot be
in a pre-merger phase.  By combining this result with: a) the
absence of any strong merging signatures in the X-ray data,
despite the relatively small separation of the two clusters, and
b) the optical light distribution mostly concentrated in the
northern clump, Gastaldello et al. (2003) suggest that the closest
core encounter between A3528-N and A3528-S happened about 1-2 Gyr
ago, with a high impact parameter. In agreement with the previous
X-ray studies (ROSAT and ASCA data, Donnelly et al. 2001), their
analysis confirms that the two main clumps in X-ray have a relaxed
structure, a centrally peaked surface brightness, and cooler
cores.

\begin{figure}
\begin{center}
{\includegraphics[width=\columnwidth]{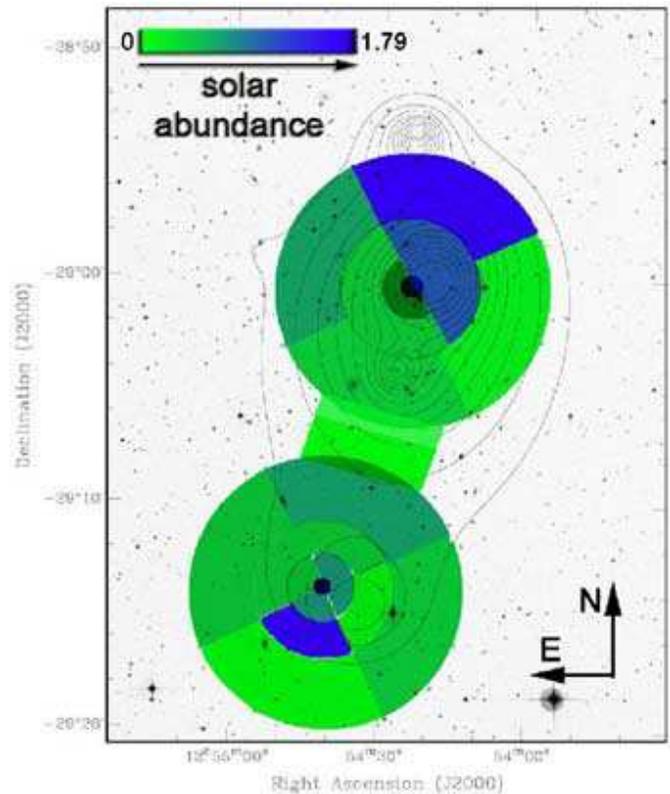}}
\caption{Iso-density contours (black) of the projected
distributions of galaxies (13$\leq$B$\leq$19.5) in the A3528
central field. The abundances in the annuli surrounding the two
main sub-clusters (A3528-N and A3528-S) and in the region between
them are shown in color (Gastaldello et al. 2003).} \label{A3528}
\end{center}
\end{figure}

\noindent The comparison of the metallicity map obtained from XMM
data (Gastaldello et al. 2003) with the results of our numerical
simulations can help to shed light on the dynamical state of
A3528. In Fig. \ref{A3528} the isodensity contours of the galaxy
number and the metallicity of the ICM in several regions are
shown. The isodensity map has been built from the APM catalogue
(13$\leq$B$\leq$19.5) on the basis of a multi-scale approach (see
Ferrari et al. 2005 for more details). The highest abundances are
observed in the SE of the core of A3528-S and in the NW region of
A3528-N, while a lower metallicity characterises the region
between the two clumps. Following our simulations, these results
would suggest that A3528-N and A3528-S are in reality in a
pre-merger phase, with a collision axis along the SE/NW direction,
in agreement with the first hypothesis about the dynamical state
of A3528 (Bardelli et al. 2001; Baldi et al. 2001; Donnelly et al.
2001; Venturi et al. 2001). It is however clear that, due to the
large errors in the abundance determination (Gastaldello et al.
2003), a higher statistics in the iron line would be useful to
confirm the pre-merger scenario.

\subsection{Metal enrichment before z=1}

\begin{figure}
\begin{center}
{\includegraphics[width=\columnwidth]{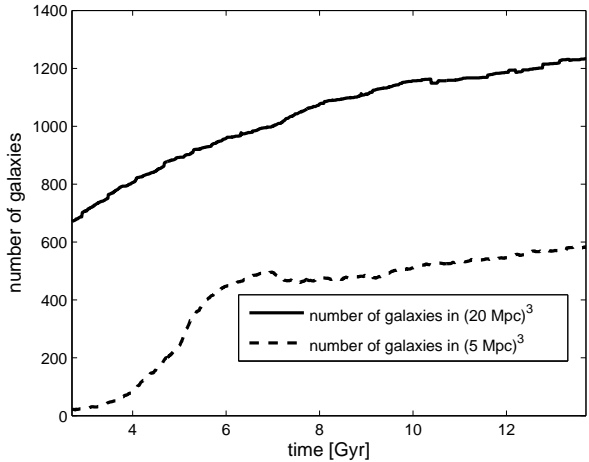}}
\caption{Evolution of the number of galaxies in the (20 Mpc)$^3$
and (5 Mpc)$^3$ simulation volume starting from z=4 ($\sim$ 1.5
Gyr) and ending at z=0 ($~sim$ 13.8 Gyr) for model cluster C as an
example is shown.} \label{galaxies113}
\end{center}
\end{figure}

As we start simulating the ICM of our model clusters at z=1, the
question arises what the fractions of metals ejected by galactic
winds and starbursts are in earlier epochs. We adress this
question by applying our wind and starburst routines to the
semi-numerical galaxies in the redshift range of z=4 to z=0. As we
do not have the knowledge of the ICM properties before z=1, we do
not suppress quiet galactic winds by pressure of the ICM on
galaxies in the central part of the galaxy cluster. In addition we
sample the ejected metals along the trajectories of the galaxies
on a (20 Mpc)$^3$ co-moving grid to get an idea of the
metal-distribution from the early epochs. The metals are only put
into cells at the positions of the galaxies without assigning any
velocities. In this very simple approach we sample all the metals
on a $128^3$ grid. This procedure helps us to see where galaxies
lose their metals, in the centre or in the outskirts. In Fig.
\ref{galaxies113} the evolution of the number of galaxies in the
(20 Mpc)$^3$ and (5 Mpc)$^3$ co-moving simulation volume starting
from z=4 for model cluster C is shown. As the cluster forms around
z=1 ($\sim$ 5.8 Gyr), the increase of galaxies in the central (5
Mpc)$^3$ region of the cluster is clearly visible.

\begin{figure}
\begin{center}
{\includegraphics[width=\columnwidth]{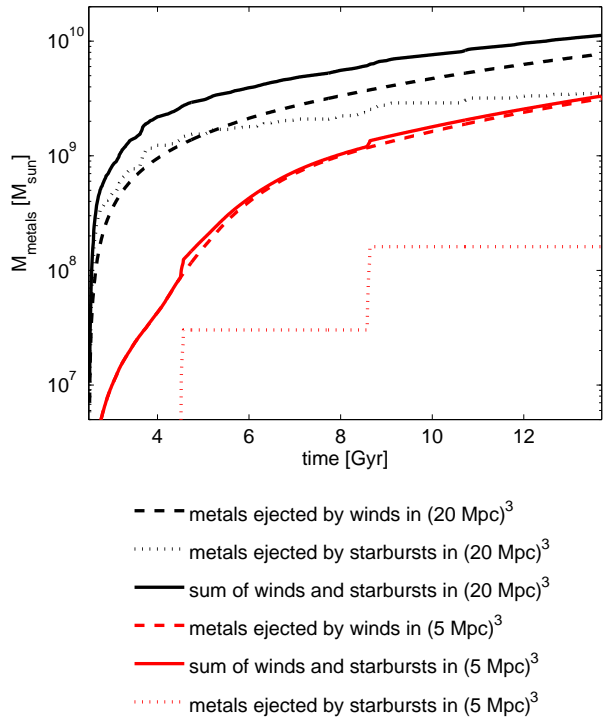}}
\caption{Amount of metals ejected into the ICM by galactic winds
and starbursts starting at z=4 for model cluster C. The black
lines show the amount in the (20 Mpc)$^3h^{-1}$  and the red lines
in the (5 Mpc)$^3h^{-1}$ simulation co-moving volume.}
\label{metalevo113}
\end{center}
\end{figure}

\noindent The amount of metals ejected by all galaxies having a
quiet galactic wind or a starburst as a function of time is shown
in Fig. \ref{metalevo113}. The black lines give the amount of
metals in the (20 Mpc)$^3$ simulation volume, whereas the red
lines give the same quantity in the (5 Mpc)$^3$ volume. The
different processes (galactic winds and starbursts) are plotted
separately. Starbursts happen more often in the outskirts of the
galaxy cluster, where the galaxies move in groups towards the
cluster. Before z=1 merger-driven starbursts are the dominating
enrichment process in comparison to galactic winds in model
cluster C. In the smaller volume, representing the cluster, only
two strong merger-driven starbursts happen since z=4. Therefore
winds dominate as enrichment process in the (5 Mpc)$^3$ volume.

\noindent The amount of metals ejected by galactic winds and
starbursts before and after z=1 is given Table
\ref{metals_before_after}. The first trend is that starbursts
occur more often in the outskirts of the cluster, but they do not
play an important role for enriching the central (5 Mpc)$^3h^{-1}$
region. Note that in this approximation the ejected gas does not
follow the potential (time dependent) as we do it in the
hydrodynamical approach derived above. This is a good
approximation as the infalling gas does not fall right to the
dense centre, but it is stopped by the ICM at large radii.
Therefore enriched outer regions (above 5-7 Mpc distance from the
cluster centre) will not have entered the cluster centre yet. Most
starbursts before z=1 happen in these outer regions and therefore
they contribute only little to the overall metal abundance in the
central cluster region (r $<$ 3 Mpc). As the number of galaxies
with mass losses by galactic winds increases with a higher rate in
the (5 Mpc)$^3h^{-1}$ volume than in the (20 Mpc)$^3h^{-1}$, see
e.g. Fig. \ref{galaxies113}, the amount of metals ejected by quiet
galactic winds in the redshift interval z=1 to z=0 is higher in
the (5 Mpc)$^3h^{-1}$ volume. This trend is even more pronounced
in the merging cluster. Galactic winds are dominating the
enrichment in comparison to starbursts. But this has an intrinsic
uncertainty, i.e. the metallicity of the starburst ejecta,
especially if the metallicity of the ejected gas in high redshift
starbursts is higher than in the semi-numerical model.

\begin{table*}
\begin{center}
\caption[]{Amount and fractions of metals ejected in the redshift
intervals z$\in[4,1]$ and z$\in[1,0]$}
\begin{tabular}{c c c c c c c}
\hline \hline & & all metals & ejected by & ejected by & ejected
by & ejected by \cr co-moving & model & ejected z$\in[4,0]$ &
galactic winds & galactic winds & starbursts & starbursts \cr
volume & cluster & [10$^9$ M$_{\sun}$]& z$\in[4,1]$ & z$\in[1,0]$
& z$\in[4,1]$ & z$\in[1,0]$ \cr \hline (20 Mpc)$^3h^{-1}$ & A &
14.2 & 29\% & 48\% & 14\% & 9\%\cr (20 Mpc)$^3h^{-1}$ & B & 14.7 &
20\% & 42\% & 17\% & 21\%\cr (20 Mpc)$^3h^{-1}$ & C & 11.3
 & 18\% & 51\% & 16\% & 15\%\cr\hline (5 Mpc)$^3h^{-1}$ & A & 5.2 & 24\% & 75\% & 0\% & 1\%\cr
 (5 Mpc)$^3h^{-1}$ & B & 2.7 & 13\% & 87\% & 0\% & 0\%\cr (5 Mpc)$^3h^{-1}$
 & C & 3.3 & 11\% & 84\% & 1\% & 4\%\cr\hline
\end{tabular}
\label{metals_before_after}
\end{center}
\end{table*}

\begin{figure}
\begin{center}
{\includegraphics[width=6cm]{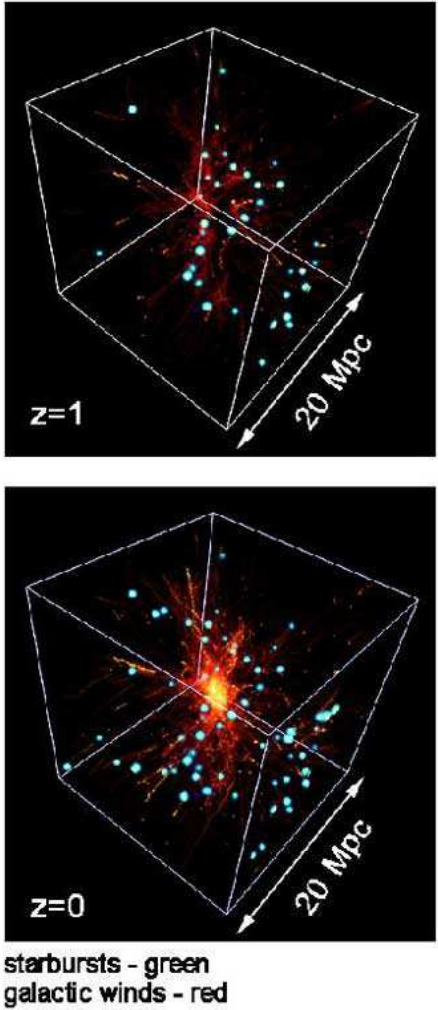}} \caption{The
distribution of metals in the (20 Mpc)$^3$ simulation volume at
z=1 and and at z=0. The amount of metals is calculated as
described in the text, without the hydrodynamic calculation of the
ICM. The green colour shows the metals ejected by starbursts,
whereas the red colour indicates metals ejected by galactic
winds.} \label{z3D}
\end{center}
\end{figure}

\begin{figure}
\begin{center}
{\includegraphics[width=6cm]{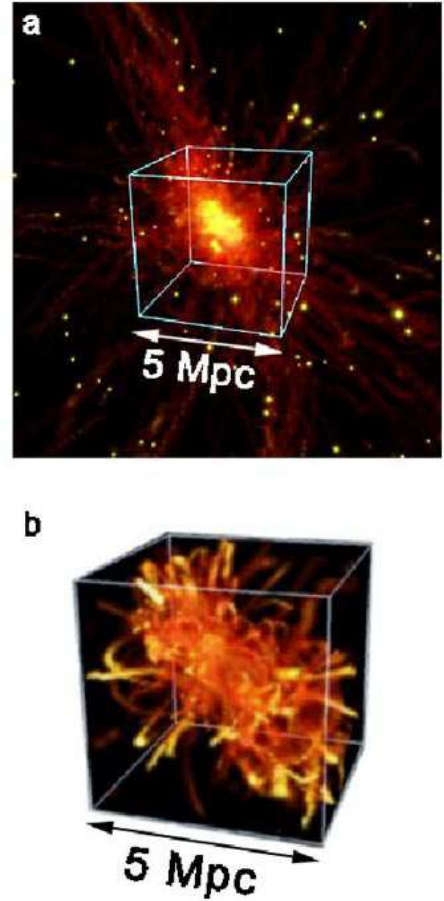}} \caption{Distribution
of metals for model cluster A at z=0. Image a shows the result of
metal ejection by galactic winds and starbursts with the simple
approach used in this section. Image b gives the same quantity
from the combined N-body/hydrodynamic simulation.}
\label{kapf_hydro}
\end{center}
\end{figure}

\noindent In Fig. \ref{z3D} the distribution of metals in the (20
Mpc)$^3$ simulation volume at z=1 and and at z=0 is shown. The
amount of metals is calculated as described before, without the
hydrodynamic calculation of the ICM. The distribution of
starbursts in the outskirts of the cluster is clearly visible,
whereas the galactic winds are the dominant enrichment process in
the central regions. At z=1 the model cluster is starting to form.
Therefore not many galaxies, with a galactic wind, are in the
cluster region. At the end of the simulation at z=0 the picture
changes: winds have enriched the central region.
\noindent In Fig. \ref{kapf_hydro} we compare the metal
distribution calculated with the simple approach without ICM of
this section and with our combined N-body/hydrodynmic approach.
The distribution is given at z=0 for model cluster A, the most
massive one. In the simple approach the metals are peaked around
the centre and not as widely distributed as in the
N-body/hydrodynmic approach (see images (a) and (b) in Fig.
\ref{kapf_hydro}). The pressure of the ICM on the galaxies in the
very central region ($\sim$ 1 Mpc around the cluster centre) does
suppress galactic winds (image b). This is not the case in the
simple approach (image a), where the metals from galactic winds
are peaked around the centre. Although the approach in this
section is very simple, the elongated metal stripes can be found
in both approaches.
\noindent Note that the mass resolution of the N-body simulation
has a limit of 1$\times10^{11}$ M$_{\sun}$ (baryonic and
non-baryonic). Galactic winds of spiral galaxies seem not to be
responsible for all of the observed metallicities in the ICM. In
the case of starbursts the mean uncertainty in our model is the
metallicity of the ejected gas, especially at high redshifts.
Although we do not resolve dwarf spiral galaxies, which will
contribute to the overall enrichment of the ICM, the uncertainty
in the metallicities in the outflows of high redshift starbursts
can alter the results most. Therefore the estimate given here
represents a lower limit.\\
There are some differences of our approach compared to other work
in this field (e.g. de Lucia et al. 2004, Aguirre et al. 2005,
Springel \& Hernquist 2003, Tornatore et al. 2004). In de Lucia et
al. (2004) (N-body simulations with a semi-analytical galaxy
evolution model) there are three transport processes modelled: the
ejection, retention and wind model. Whereas the wind model
enriches directly the ICM, the retention model enriches the halo
of the galaxy (cooling is included) and the ejection model
enriches the ICM with metals, which are then reincorporated into
the host halo in a given time interval. Their result is that the
ejection model contributes the highest fraction of metals to the
ICM at present epochs. The observed metallicities in the ICM
(taken from X-ray observations) are compared with the sum of
metals in the halos of galaxies and the ICM in their simulation.
Our wind model takes into account only gas which is not
gravitationally bound to the galaxy furthermore and is therefore
heated by the ICM to temperatures, in which X-rays are emitted in
an energy range accessible to current X-ray telescopes. Although
supernovae driven outflow enrich the hot gas in the halo of a
galaxy, only a minor fraction of enriched gas will be ejected into
the ICM, if the SFRs are low (about 1 M$_{\sun}$/yr). The major
fraction will not reach the escape velocity, will cool radiatively
cool and become then a part of the cold disk gas again.\\
In Aguirre et al. (2005) combined N-body/SPH simulations were done
to confront cosmological simulations with observations of
intergalactic metals. Their applied wind model and cooling
description is not able to reproduce the observed amount of metals
in the intergalactic medium (IGM). The metallicity map given in
Springel \& Herquist (2003) does not distinguish between hot IGM
gas and gas in the halo of the galaxies. Tornatore et al. (2005),
combined N-body/SPH simulations on galaxy cluster scales and give
profiles of the Fe abundance in simulated galaxy clusters.
Although the profiles do not match the spatial distribution of
metals observed in galaxy clusters, the overall amount of metals
seems to be reproduced. Unfortunately X-ray emission weighted
metal maps are not given to
see, if the metals would be observable in X-ray observations.\\
A major difference in our model compared to de Lucia et al.
(2004), Springel \& Hernquist (2003), Aguirre (2005) and Tornatore
et al. (2004) is the calculation of X-ray emission weighted metal
maps. As we are interested in a direct comparison with X-ray
observations we treat the ICM hydrodynamically (including
radiative cooling (Sutherland \& Dopita, 1993)) and extract
artificial X-ray maps from our simulations to compare them
directly with real X-ray observations of galaxy clusters. This
allows us to conclude that galactic winds and starbursts are in
general not sufficient to enrich the ICM of galaxy clusters up to
about 0.3 solar abundance in metallicity, which is observed in
galaxy clusters in the X-ray regime. Additional processes like
ram-pressure stripping (see Domainko et al. 2005), stripping of
hot halo gas or jets of AGN are not taken into account in this
work. Their contribution is addressed in Domainko et al. 2005 and
subsequent papers.

\subsection{X-ray weighted metal maps of our model clusters as
            observed with Chandra, XMM-Newton and XEUS}

\begin{figure}
\begin{center}
{\includegraphics[width=\columnwidth]{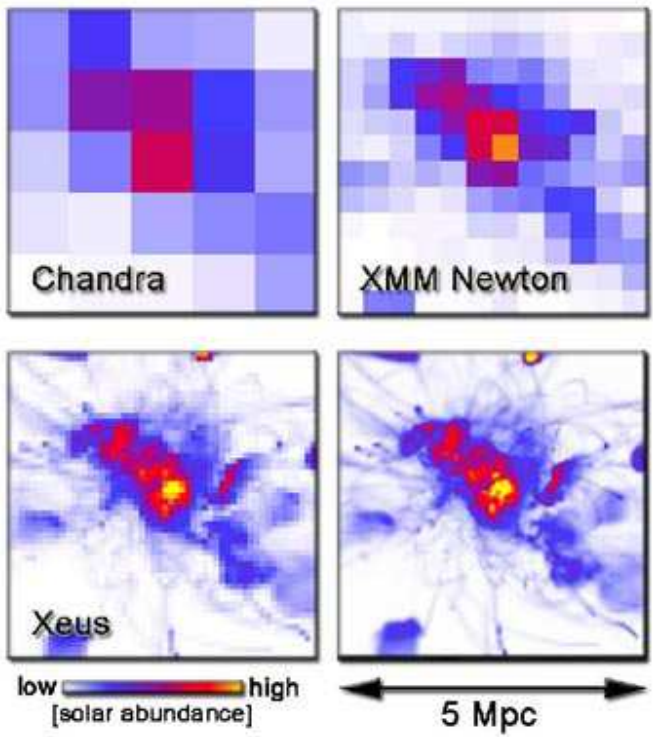}}
\caption{X-ray weighted metal maps as they would be observed with
Chandra, XMM-Newton and the next generation X-ray telescope XEUS
for a given exposure time. Lower right image: the real metal map.
Different sensitivity of the different missions yield different
resolutions of the metal map.} \label{xeus}
\end{center}
\end{figure}

As mentioned above, observations are always limited by the number
of X-ray photons. The X-ray weighted metal maps obtained from our
simulations have resolutions which cannot be reached by state of
the art observations with instruments like Chandra or XMM-Newton.
In Fig. \ref{xeus} four X-ray weighted metal maps of model cluster
B at z=0 are shown. Each map has a different resolution,
reflecting the sensitivity of different X-ray instruments. The
last map has the resolution of the simulation. Assuming that a
cluster resides at a given distance which enables us to extract a
5$\times$5 pixel X-ray weighted metal map with the X-ray telescope
Chandra, XMM-Newton would be able to obtain a 12$\times$12 pixel
metal map within the same exposure time. This is due to the fact
that XMM-Newton has a six times higher sensitivity than Chandra.
XEUS is planned to be a follow-up mission of XMM-Newton, which
will have about a sensitivity 20 times higher than XMM-Newton. For
diffuse low surface brightness objects, like galaxy clusters, the
effective area is a major issue, therefore for a reasonable
observation time, Chandra maps will have lower resolution than
XMM-Newton. With XEUS we will be able to resolve for the first
time the metal distribution in the ICM on the scales of single
galaxies in nearby clusters.

\subsection{On the survival timescales of inhomogeneities in the metallicity distribution}

\begin{figure}
\begin{center}
{\includegraphics[width=\columnwidth]{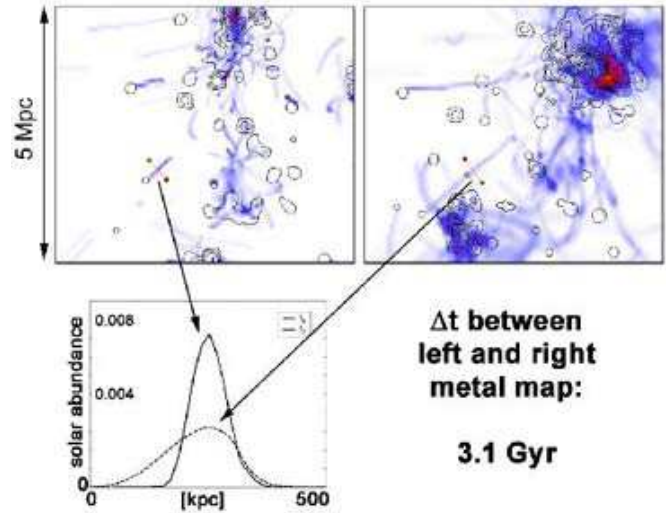}} \caption{Two
metal maps of model cluster B for two different times, 3.1 Gyr of
evolution between them. Along the red line the metal distribution
is integrated and plotted.} \label{metalsurvive1}
\end{center}
\end{figure}

As shown in the X-ray weighted metal maps, metallicity
inhomogeneities can survive over long periods. The timescales
depend strongly on the environment of the enriched regions in the
ICM. If a highly enriched region is situated in the central part
of a merging cluster, the metals will be spread quickly over huge
volumes, due to the internal kinematics of the ICM. Perturbations
propagating from the central regions into the outer parts blur
enriched regions on shorter timescales, than in the outskirts (the
low dense environment of a galaxy cluster). To give an upper limit
of the survival timescale of these inhomogeneities, we
investigated a stripe of enriched ICM, caused by a galaxy having a
quiet galactic wind. In Fig. \ref{metalsurvive1} two X-ray
weighted metal maps for model cluster B are given, with a time
difference of 3.1 Gyr. In the metal maps we measured a one
dimensional metal distribution highlighted by two lines. Below the
two metal maps the metallicity distribution as a function of space
is shown for the two timesteps. As the enriched region diffuses,
the maximum decreases by more than 50\%. If we integrate over the
2D distributions and compare them, the overall metallicity
decreases in this region by a factor of 0.12. The explanation for
this decrease in the metallicity is the increase of density in the
same region in the same time interval by a factor of 0.14. The not
previously enriched ICM coming from other mergers partly mixes
with the enriched region.
\noindent Of course this effect depends strongly on the
environment. In the innermost parts of a galaxy cluster metals are
mixed with the ICM on shorter timescales, i.e. within several 100
Myr, whereas these inhomogeneities should be observable in the
outer parts over several Gyr. In strongly merging galaxy clusters
the timescales should be shorter than in relaxed massive clusters.
This trend is noticeable in our simulations as well.

\section{Summary and Conclusions}

We use combined N-body/hydrodynamic calculations with
semi-numerical galaxy modelling to investigate galactic winds and
starbursts as an enrichment process of the intra-cluster medium
(ICM). The mass-loss rates by quiet galactic winds for spiral
galaxies are calculated with a code developed by Breitschwerdt et
al. (1991). The terminus mass loss rate in our simulations means
galactic matter, which is not gravitationally bound to the
originating galaxy anymore. As this enriched matter is then part
of the ICM, we include it into our hydrodynamic ICM simulations.
We then investigate the spatial distribution of the ejected matter
and give X-ray surface brightness, X-ray emission weighted
temperature and metal maps to study the dynamical state of the
model galaxy clusters. We find that

\begin{itemize}

\item{there is a difference between the strength of the enrichment
by galactic winds and starbursts in merging and in non-merging
model clusters. Merging systems show more galaxy-galaxy
interactions, which cause more starbursts. On the other hand
galactic winds can be suppressed by the pressure of the ICM onto
the galaxies in relaxed clusters. As relaxed systems have more ICM
to enrich, galactic winds are less effective. In our simulations
we find that galactic winds and starbursts can contribute up to
5\% (merging cluster) to the observed metallicities in galaxy
clusters in the redshift interval from z=1 to z=0 and up to 15\%
in the redshift interval from z=4 to z=1.}

\item{X-ray surface brightness,  X-ray emission weighted
temperature and metal maps can give a deeper insight into the
dynamical state of a galaxy cluster. Temperature and metal maps
are harder to obtain in observations (limited by the number of
X-ray photons), but they yield a lot of information. We show that
especially the metal maps can help to distinguish between the pre-
or post merger state of a galaxy cluster. Pre-mergers have a
metallicity gap between the subclusters, post-mergers have a high
metallicity between subclusters. This holds also if the resolution
is reduced to 9x9 pixels in our artificial X-ray weighted
metallicity maps, which cover an area of (5 Mpc)$^2$.}

\item{the inhomogeneities which are introduced by galactic winds
and starbursts into the ICM can survive for a long time. In the
outskirts of a galaxy cluster, where the variation in the density
and the velocity field of the ICM is less than in the inner parts,
we find that inhomogeneities can survive up to several Gyr. The
timescales for the inhomogeneities are different for merging and
non-merging galaxy clusters, because of mixing by shocks,
introduced by major merger events (in the ICM).}

\item{metal enrichment of the ICM by galactic winds before z=1 is
not as strong as after z=1 in the cluster centre ((5
Mpc)$^3h^{-1}$ volume), whereas for merger-driven starbursts there
is no noticeable difference. Before z=1 starbursts enrich the ICM
on large scales at a comparable rate as galactic winds. As
merger-driven starbursts happen more often in the outskirts of
clusters, they enrich the outer (larger r$\sim$3 Mpc) regions of
galaxy clusters. Galactic winds, on the other hand occur
throughout the whole cluster, except for the densest innermost
regions, where they might be suppressed by the high ICM pressure.}

\end{itemize}

\noindent Note that the metal maps in this paper are directly
comparable to observed metal maps, as we only take ejected matter
into account for the enrichment processes, which would be
observable in X-rays. The observed metallicities in galaxy
clusters are one order of magnitude higher than in our
simulations. We believe that galactic winds are not able to
produce the total observed amount of metals in the ICM. Additional
processes like ram-pressure stripping, hot halo gas stripping,
jets of AGNs or intra-cluster supernovae obviously play an
important role as well. A major uncertainty are the metallicities
of the ejecta of high redshift starbursts, which will be
investigated in an upcoming paper.

\section*{Acknowledgements}

The authors are grateful to the anonymous referee for his/her
criticism that helped to improve the paper. The authors
acknowledge the Austrian Science Foundation (FWF) through grant
number P15868, UniInfrastrukturprogramm 2004 the bm:bwk
"Konsortium Hochleistungsrechnen", the bm:bwk Austrian Grid (Grid
Computing) Initiative, the Austrian Council for Research and
Technology Development and the support of the European
Commission's Research Infrastructures activity of the Stucturing
the European Research Area programme, contract number
RII3-CT-2003-506079 (HPC-Europa). In addition the authors
acknowledge the DFG grant Zi 663/6-1, the Marie Curie individual
fellowship MEIF-CT-2003-900773 (CF) and the European Commission
through grant number HPRI-CT-1999-00026 (the TRACS Program at
EPCC). Edmund Bertschinger and Rien van de Weygaert are
acknowledged for providing their constrained random field code and
Joshua Barnes and Piet Hut for their tree code. The authors would
like to thank Sabine Kreidl for many useful discussions.

\end{document}